\documentstyle[aps,pre,epsfig]{revtex}

\begin{document}

\title{Statistical mechanics of fluidized granular media:
short range velocity correlations} 

\author{R. Soto and M. Mareschal\\
{\em \small CECAM, ENS-Lyon, 46 All\'ee d'Italie, 69007 Lyon, France}}

\maketitle

\begin{abstract}
A statistical mechanical study of fluidized granular media is 
presented. Using a special energy injection mechanism, homogeneous
fluidized stationary states are obtained. Molecular dynamics
simulations and theoretical analysis of the inelastic hard-disk model
show that there is a large asymmetry in the two-particle
distribution function between pairs that approach and separate. Large
velocity correlations appear in the postcollisional states, 
due to the dissipative character
of the collision rule. These correlations can be well characterized by
a state dependent pair correlation function at contact. 
It is also found that velocity
correlations are present for pairs that are about to
collide. Particles arrive to collisions with a higher probability 
that their velocities are parallel rather that antiparallel. 
These dynamical correlations lead to a decrease of the
pressure and of the collision frequency as compared to their Enskog
values. A phenomenological modified equation of state is presented.

\end{abstract}


\section{Introduction}

There has been much interest recently devoted to the description of
granular media. The understanding of their physical properties is important
because they appear in many different phenomena taking place in our daily
life as well as in various industrial processes. Experiments have been
devised which have permitted to focus on many different and new aspects.
When subjected to injection of energy (vibrated plates for example), they
present many similarities with fluids: convection take place, patterns can
form and time-periodicity may be observed
\cite{herrmann92,aoki96,melo,jaeger,falcon99,kadanoff99,wildman99,Rosa00}.

For those fluidized granular media, careful investigations have permitted
to test the adequacy of hydrodynamical continuum equations generalized to
take into account the dissipation of energy due to the inelasticity of the
collisions between the grains. Those equations are usually closed with the
help of an equation of state and of phenomenological laws inspired by the
hard-sphere model for fluids
\cite{campbell90,jenkins8385c,deltour97,helal97,bizon99,sotomar99}.
Also, kinetic theory models have been studied obtaining good accord
with computer simulations in the low density cases
\cite{lun84,brey96a,noije98}. 

In most of these works, granular media are described microscopically by 
means of the Inelastic Hard Sphere (IHS) model. Grains are modeled as
soft hard spheres that dissipate energy at collision through a
constant restitution coefficient $\alpha$. It is quite remarkable that such
a simple model can already account for many peculiarities of the
observed behavior
\cite{goldhirsch93,mcnamara96,grossman97}.

In this paper, we would like to go one step further: namely to investigate
the IHS model as if it was a genuine statistical
mechanical model for a granular fluid, and examine the effects of the
inelasticity 
on the statistical properties of the fluid. One first obvious difference is
the absence of any equilibrium state. Let to itself, an assembly of
inelastic hard spheres will evolve towards a final state with no motion.
One can however achieve stationary states by allowing contact with an
energy reservoir, but those states are nonequilibrium states with a
permanent energy flow coming from the reservoir and dissipated at
collisions into the internal elasticity of the grains (and neglected for
our purpose). It is our aim to describe these Non Equilibrium Steady
State (NESS), considering the dissipativity $q$, defined as $q=(1-\alpha)/2$,
as the parameter responsible for deviations from the equilibrium hard
sphere system. 

Letting a system composed of IHS grains is let to evolve freely with
periodic boundary conditions, it cools down homogeneously. This
homogeneous cooling state (HCS) has been widely studied
\cite{brey96a,mcnamara96,haff83,noije98c}, and because of its
simplicity the HCS has been used as a reference  
state to build up theories for non-homogeneous states. It is the
analog of the equilibrium state (with a Maxwellian distribution)
for elastic systems \cite{Toto,brey98,garzo99}.
Among other features, it has been shown that, in this state, 
the velocity distribution function is not Maxwellian, having a
nonvanishing fourth cumulant and a long velocity tail
\cite{noije98c}. Also, long range velocity correlations are developed, 
in the form of vortex fluctuations \cite{noije97}.

It has been realized that, for any given
value of the inelasticity, whatever small,  there is always a size over
which the homogeneous reference state looses its stability. The system 
undergoes a transition 
towards a state with a shear flow spontaneously developing and possibly
towards an inhomogeneous clustered state
\cite{deltour97,goldhirsch93,mcnamara96,brey96,ShearInst}.
In order to characterize 
intrinsic properties like an equation of state, one needs to restrict to
consider stable homogeneous states before any instability occurs.
This means in general to consider slight inelasticities and
finite system sizes where $1/N$ effects are present, $N$ being the
number of grains.

Computer simulation techniques has proven very valuable in granular
studies: like in molecular dynamics (MD), computers are used to integrate
numerically the equations of motion of a few hundred to a few thousand
grains whose interaction is usually limited to inelastic collisions. The
computed behavior compares generally well with experiments, which give
confidence in the adequacy of the inelastic hard sphere model for
accounting for most of the specificities of granular fluids. We are going
to use such a technique with a mechanism added in order to keep the kinetic
energy of the grains constant.

The essential result obtained is the characterization of short range
correlations which are built in granular fluid models
and which determine both the static (equation of state) and dynamical
properties (kinetic equation) of the granular fluid. The pressure of
the stationary state is computed from its mechanical definition and we
can show that it is related to the pair correlation function at
contact for precollisional configurations only.  This induces an
extra variable in the pair correlation function, the angle between
relative distance and relative velocity at collision.

Second, the kinetic description of the stationary state is well described
by the (modified) hierarchy equations. In order to write a closed equation for
the one-particle distribution function, like the Enskog equation for hard
spheres, some assumptions are made on the two particle distribution
function. In Enskog's equation, it is assumed that the particles
that collide do not have velocity correlations (or dynamical
correlations). When two particles are at contact with approaching
velocities (precollisional state), it is assumed that the two particle
distribution function $f^{(2)}$ can be well approximated by~\cite{Lutsko}
\begin{equation}
f^{(2)}({\bf r}_1,{\bf v}_1,{\bf r}_2,{\bf v}_2)
\Theta(-{\bf v}_{12}\cdot{\bf r}_{12}) \delta(r_{12}-\sigma) =
 \chi f^{(1)}({\bf v}_1) f^{(1)}({\bf v}_2) 
\Theta(-{\bf v}_{12}\cdot{\bf r}_{12}) \delta(r_{12}-\sigma)
\label{nocorr}
\end{equation}
where $\chi\!=\!g(\sigma^{+})$ is the pair correlation function at
contact and the Heaviside function guarantees that it is a
precollisional state. Note that in elastic systems at equilibrium, the
previous relation is exact.  
We show here that this assumption is no more true for
inelastic hard spheres, with an increased probability that colliding disks or
spheres are parallel rather than antiparallel, at least for moderately
dense fluids. 
Such velocity correlations might also appear in elastic fluids but
only under nonequilibrium  regimes (typically over macroscopic
scales). Here they appear on the shortest scales, those which
correspond to molecular distances, only because of the dissipative
character of the microscopic dynamics.

The article is organized as follows: first, we present with some detail the
simulation technique which has been used. As already mentioned, we
simulate grain dynamics at constant energy: at every collision, all
particle velocities are rescaled so as to keep the total kinetic energy
constant. We present the technique and emphasize in particular its
equivalence with a time-rescaling change. Next, we analyze the pressure and
the pair-correlation function in the stationary state by making time
averages of observables and configurations. The analysis lead in particular
to the need of considering velocity correlations which, as will be shown,
are also responsible for a pressure drop. This is done in the fourth
section, before the conclusions.

\section{Constant energy simulations of the IHS model} \label{SimTechinque}

In the HCS the system is in a non-stationary
state that does not allow to make time averages of different
properties. It is then necessary to inject energy to the system in
order to keep is stationary. Different methods have been used, mainly the
injection of energy trough the walls (vibrating or stochastic), by
external fields (like the flow in a pipe) or by 
stochastic forces acting on the particles
(see for example \cite{williams96b}). These  
methods have the disadvantage of destroying the homogeneity of the
fluid or, in the case of the stochastic forces, adding other dynamical 
effects \cite{ErnstPrivate}.

In this article we use a thermalizing method that both preserves the
homogeneity and the dynamical properties of the granular
fluid. Formally, each time two particles collide, the
dissipated energy in reinjected immediately into the system by
multiplying all particle velocities by the same factor. This factor is 
chosen each time in order to keep kinetic energy constant.
As the IHS model does not have an intrinsic energy (or time) scale the 
rescaling of all velocities does not change the evolution of
the system, the collision sequence (thus all physical phenomena) is
preserved but viewed at a different speed. In this sense this
thermostating mechanism is the most appropriate for the HCS.

In this nonequilibrium steady state (NESS) all macroscopic quantities
are stationary and can be averaged in numerical simulations.
If we set the initial energy to give
a temperature equal to one, then all averaged quantities correspond  to 
computing them at this temperature. To obtain the value at another 
temperature, simple dimensional analysis gives the desired value. For example,
if $p_{\rm NESS}$ is the computed pressure in the NESS, the actual
value of the pressure at another temperature $T$ is $p=p_{\rm NESS} T$.

The IHS model is simulated using event driven molecular dynamics
\cite{MRC}. A direct implementation of the thermalizing method would
lead to a computational cost proportional to the total number of
particles for each collision, making impossible to make long
simulations. But we can take advantage of the lack of an energy scale 
to make it much faster. In effect, rescaling all velocities is
equivalent to rescaling the time. We define a rescaled time by
the relation $ds = \gamma\, dt$ with
\begin{equation}
\gamma = \sqrt{\frac{E(t)}{E(0)}} \label{gamma}
\end{equation}
where $E$ is the total energy in the system. Note that $\gamma$ is a
piecewise constant decreasing function. It can directly checked
that if rescaled velocities are defined as
$\widetilde{{\bf v}}=d{\bf r}/ds$, then the rescaled kinetic energy
($\widetilde{K}=\sum \frac{m}{2} \widetilde{v}_i^2$) is conserved. 
This transformation have been successfully used to study the shearing 
instability~\cite{ShearInst}.

Then, the simulation is done for the cooling IHS model (that is, no
velocity rescaling is done) but at each collision the new
kinetic energy is computed and $\gamma$ is evaluated using Eq.~\ref{gamma}. 
In the simulation, then, the energy and the
average velocity decrease, but the rescaled velocities are computed using
$\widetilde{{\bf v}}= {\bf v}/\gamma$. Having the rescaled velocities, all the 
static properties can be computed directly. 
Since the function $\gamma$ is  piecewise constant, the $s$-time can be 
integrated in the simulation; this allows to make periodical
measurements in the NESS (equally spaced in $s$).

Finally, to avoid roundoff errors, each time the kinetic energy has decreased 
by some orders of magnitude
(typically $10^{-7}$ of the initial value) a real velocity rescaling is
done to put again the temperature equal to one. In this process also the 
center of mass velocity is subtracted because, otherwise, small errors will 
be amplified by the rescaling.

As already mentioned the HCS becomes unstable (shearing instability)
for certain values of dissipativity, density, and system size. Given
values for the dissipativity and density, the system size (number of
particles) is constrained to be smaller than a certain value to keep
the system homogeneous. In 2D for a low density and low dissipation
case, it is given by \cite{mcnamara96}
\begin{equation}
N_{max} = \frac{\pi}{q n}
\end{equation}

For larger densities there are more complex expressions written in
terms of the transport coefficients, but in all cases there 
is a maximum system size over which the system becomes unstable.
This phenomenon does not allow to simulate large systems (for example,
for $n=0.2$ and $q=0.02$, $N_{max}=800$) and finite size effects are
obtained. To avoid these effects, simulations with
different number of particles are done and then the results are
extrapolated to the infinite size limit. The extrapolation is done
using the standard model for any quantity that does not vanish in this
limit, that is, for any quantity $A$ its size dependence is modeled as
\begin{equation}
A(N) = A_{\infty} + A_1/N \label{extrapola}
\end{equation}

When $N_{max}$ is smaller than around 1000, the effects of the shearing
instability are present before the critical value as fluctuations
in the $k=2\pi/L$ mode of the shearing velocity
\cite{ShearInst}. These fluctuations 
have both large amplitudes and long correlation times (divergent at
the critical point). Some quantities couple strongly with this
fluctuating field (for example, the pressure) and small
system sizes must be studied to extrapolate to infinity.

We present series of simulations in 2D for the IHS model in the non
equilibrium steady state already described. Dimensions 
are chosen such that the disk diameter, particle masses and initial
temperature are set equal to one. 
The simulations are done at three different number densities ($n=N/V$):
$n=0.05$, $n=0.1$, and $n=0.2$. In each case the dissipation
coefficient takes the values $q=0$, $q=0.002$, $q=0.005$, $q=0.01$, and 
$q=0.02$. Given the
small values of the dissipation coefficients, long simulations are
needed to obtain good statistics in order to identify the
$q$-dependence of the different quantities.

As the simulations are done for small dissipations, the results are
presented as series in the coefficient $q$. Also, when possible, the
results are condensed in power expansions in the density, but these
expressions must not be interpreted as assuming that these quantities
are necessarily analytic in density or dissipativity.

Finally, unit s are chosen such that the disk diameter $\sigma$ and
the particles masses are set to one. Also, the temperature in the NESS 
is fixed to one.


\section{Virial Pressure}

As in granular media there is no equivalent to a free energy or an
entropy, the pressure can not be defined thermodynamically but only
mechanically.
We use the virial expression for the pressure \cite{Hansen} that is a
mechanical definition valid for any isotropic system composed
of particles that interact with pairwise forces. If the total volume
is $V$ and the total kinetic energy is $K$, in 2D the pressure is
given by
\begin{equation}
p = \frac{\left<K\right>}{V} - \frac{1}{4V}\left<\sum_{i\neq j}
{\bf r}_{ij}\cdot {\bf F}_{ij} \right> \label{pvirial1}
\end{equation}

In the NESS, kinetic energy is constant and equal to $N$ 
and the forces are impulsive ones. Using the collision rule, it is
easy to check that the forces are given by
\begin{equation}
{\bf F}_{ij} = -(1-q) \left|{\bf v}_{ij}\cdot\hat{r}_{ij}\right|
\hat{r}_{ij}\, \delta(t-t_{ij})  
\end{equation}
where $t_{ij}$ is the collision time for the pair $i\!-\!j$, and the
relative velocity is evaluated before the collision.

Then, if the average is written as a time average, the delta functions 
are integrated giving
\begin{equation}
p = n + \frac{1-q}{4V \tau} \sum_{\mbox{colls}}
\left|{\bf v}_{ij}\cdot{\bf r}_{ij}\right| 
\end{equation}
where $\tau$ is the averaging time, and it has been used that particles 
diameter is set to one.

In simulations we measure the quantity
\begin{equation}
p_1 = \frac{1-q}{ N_{\mbox{colls}}} \sum_{\mbox{colls}}
       \left|{\bf v}_{ij}\cdot{\bf r}_{ij}\right|\\
\end{equation}
in terms of which the total pressure is given by
\begin{equation}
p = n +\frac{n \nu}{4} p_1 \label{pvirial2}
\end{equation}
where $\nu$ is the collision frequency.

The results from the simulations, fitted linearly with $q$ are
\begin{eqnarray}
p_1(n=0.05) &=& 1.772 (1-q) - 0.163 q \\
p_1(n=0.1)  &=& 1.772 (1-q) - 0.262 q \nonumber\\
p_1(n=0.2)  &=& 1.772 (1-q) - 0.528 q \nonumber
\end{eqnarray}

That collected can be fitted to
\begin{equation}
p_1 = \sqrt{\pi} (1-q) -2.67\, q\, n \label{p1num}
\end{equation}

The average in Eq.~\ref{pvirial1} can also be written also as an ensemble
average

\begin{eqnarray}
p &=& n + 
\frac{(1-q)}{4V} 
\int f^{(2)}(1,2)
\left|{\bf v}_{12}\cdot{\bf r}_{12}\right|^2 \nonumber\\
&&
\Theta(-{\bf v}_{12}\cdot{\bf r}_{12})
\,d{\bf r}_1\, d{\bf v}_1\, d{\bf v}_2\, d\theta
\end{eqnarray}

\noindent where $\theta$ is the angle between the relative velocity and the
relative position.
Assuming lack of velocity correlations at collisions, the 
two particle distribution function can be replaced by Eq.
\ref{nocorr}. In this case, the integrals can be done explicitly giving
\begin{equation}
p = n + \frac{n^2 \pi (1-q) \chi}{2}
\end{equation}
Then, in this approximation, $p_1$ can be expressed as
\begin{equation}
p_1 =  \frac{2 n \pi \chi}{\nu} (1-q) \label{p1nocorr}
\end{equation}

To study the validity of this approximation and compare with the
numerical results~(\ref{p1num}), we need to study the pair
correlation function at contact, $\chi$, and the collision frequency,
$\nu$.

\subsection{Pair correlation function at contact}

For hard particles systems, 
the pair correlation function at contact $\chi$ is defined in elastic
systems as $\chi=g(r\!\!=\!\!\sigma^{+})$, where $g(r)$ is the pair correlation
function and $\sigma$ is the particle diameter.
In granular media, this definition is somewhat ambiguous
and a direct application of the classical computational methods \cite{Allen} to
obtain $\chi$ does not give the most relevant result.

The pair correlation function is defined in elastic systems as the
probability of having two particles separated at a distance $r$, with an
adequate normalization. This definition does not take into account the
relative motion of the two particles since it is known that in equilibrium,
positions are not correlated with velocities. But, as it is shown bellow, in
granular media positions and velocities are highly correlated even in the
low density limit, having the pair correlation function different behavior
if the two particles approach or separate.

In Enskog's kinetic theory, the $\chi$ factor that appears in
Eq.~\ref{nocorr}, in the
collisional term, in the virial pressure, and in the transport
coefficients must be understood as the pair correlation function at contact
for particles that are approaching and not for the ones that separate.
In what follows we define and describe the basic properties of a
generalized $\chi$ coefficient that depends on the dynamical state of
the particles. 

We define the pair correlation function $g(r,\theta)$ as proportional
to the number of pairs that are separated by a distance $r$ and there
is an angle $\theta$ between the relative position and the relative
velocity. The normalization is chosen
such that $g$ goes to one for 
large distances. This function is computed in an analogous way as the
usual pair correlation function \cite{Allen}. The generalized pair
correlation function at contact $\chi(\theta)$ is defined as
$\chi(\theta)=g(r=\sigma^{+},\theta)$.

It can be identified two classes of pairs according to the value of the 
angle $\theta$: if $\cos(\theta)$ is positive then the particles are
separating (postcollisional states), and if it is negative the two
particles approach (precollisional states).
In the Appendix~\ref{apendix.chi} we show that the postcollisional
part of $\chi(\theta)$ can be expressed in terms
of the precollisional one . Using the collision rule and the
conservation of probability during a collision it is
found that if $\theta<\pi/2$ (postcollisional)
\begin{equation}
\chi(\theta) = \left[\cos(\theta)^2+ \alpha^2 \sin(\theta)^2 \right]^{-1}
\chi\left[\pi-\tan^{-1}(\alpha \tan(\theta)) \right] 
\label{chipostpre}
\end{equation}
where $\left[\pi-\tan^{-1}(\alpha \tan(\theta))\right]>\pi/2$ is the
precollisional angle that gives the postcollisional angle $\theta$.

In the Enskog approximation, the precollisional pair correlation
function is a constant $\chi_0$. In this approximation, the complete
function is
\begin{equation}
\chi(\theta) = \left\{ \begin{array}{ll}
 \chi_0 & \cos(\theta) \leq 0\\
 \chi_0\left[\cos(\theta)^2+ \alpha^2 \sin(\theta)^2 \right]^{-1} &
 \cos(\theta)>0 
\end{array}\right. \label{chienskog}
\end{equation}

This function is discontinuous at $\theta=\pm \pi/2$. 
Its average is $\chi_0\frac{1}{2}\left(1+\frac{1}{\alpha} \right)$,
that can be understood knowing that the postcollisional normal
relative velocity is $\alpha$ times smaller than the precollisional
one. That makes that the colliding pair rests $1/\alpha$
times longer in the postcollisional position, giving the previously
obtained average (formally the time the
particles stay at contact is zero, but $\chi$ is the limit of $g(r)$
that can be well defined for finite bins where the pair stays
finite times).

This kind of discontinuity in $\chi(\theta)$ has been found in sheared
elastic fluids 
\cite{Lutsko}. In that case the origin of the discontinuity is the
nonequilibrium character of the one-particle distribution function. In
our case it is originated by the non-conservative collision rule.

\begin{figure}[htb]
\begin{center}
\epsfclipon
\epsfig{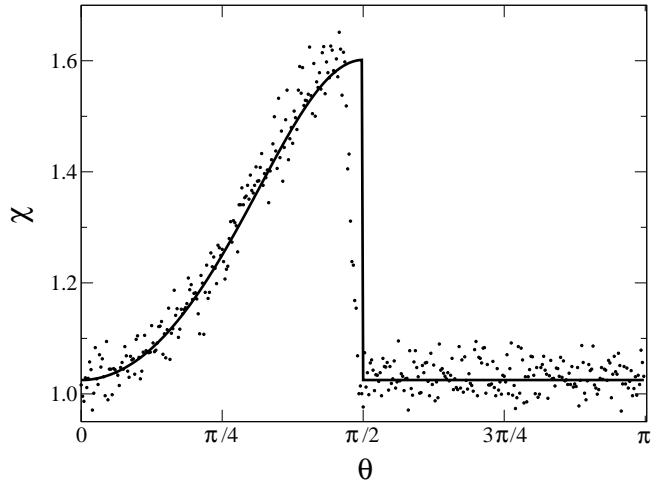} 
\end{center}
\caption{Estimation of 
$\chi(\theta)$ obtained in MD simulations (dots) compared to
the theoretical prediction. The simulation parameters are $N=1000$,
$n=0.02$, and $q=0.1$. The discrepancy near $\theta=\pi/2$ is explained in the
text.}
\label{chi.fig}
\end{figure}

The discontinuity in $\chi(\theta)$ makes that the extrapolation of
$g(r,\theta)$ to contact presents numerical problems. For the
precollisional angles $g$ is a smooth function but for the
postcollisional ones there is a discontinuity line arriving to
$\theta=\pi/2$, that prevents from obtaining good estimations of $\chi$ 
for angles slightly below $\pi/2$.
In Fig.~\ref{chi.fig} it is presented a numerical estimation of 
$\chi(\theta)$ obtained in MD simulations for a low density case.
 The comparison with the Enskog theoretical value
(Eq.~\ref{chienskog}) is good
except for a $80^\circ<\theta<90^\circ$ where is was expected to fail.

In the clustering regime of large systems (where the system is no more
homogeneous) it has been observed a large dependence of $\chi$ on
$\theta$, for the precollisional angles \cite{luding98e,BritoPrivate}.
In our case, statistical errors in the results of molecular dynamics
simulations do not allow to determine if there is a dependence for
precollisional angles, contrary to the clear dependence
for the postcollisional angles. Nevertheless if $\chi$ depends on the
precollisional angle, it is small and not like the one reported in
the clustering regime.

We define the {\em pair correlation function at contact}, $\chi$, as
the average of $\chi(\theta)$ over the precollisional angles only. 
The simulation results for $\chi$, fitted  linearly with $q$, are

\begin{eqnarray}
\chi(n=0.05) & = & 1 + 0.0655 + 0.051 q \label{chiMD}\\
\chi(n=0.1)  & = & 1 + 0.1389 + 0.053 q \nonumber\\
\chi(n=0.2)  & = & 1 + 0.3129 + 0.070 q \nonumber
\end{eqnarray}

An empirical expression for $\chi$ in the 2D elastic case is \cite{Henderson}. 
\begin{equation}
\chi(q=0) = 1+ \frac{\pi (25 - 4 n \pi)}{4(4- n \pi)^2} n
\end{equation}

A comparison with the measured values show that, for the IHS model,
there is  a very small dependence on $q$. Within the statistical error
it can be said that $\chi$ does not depend on $q$ and its value is the
same as in the elastic case.

\subsection{Velocity distribution and collision frequency}

It is known that the velocity distribution function for the HCS of the 
IHS model is not a Maxwellian but a distorted one. In the low velocity 
region, it is predicted that the fourth cumulant is not zero and
using the Boltzmann-Enskog equation it has been predicted its value
\cite{noije98c}. The fourth cumulant is defined as
\begin{equation}
k_4= \frac{\left<v^4\right>
  - 2\left<v^2\right>^2}{\left<v^2\right>^2}
\end{equation}
and the predicted value using the Boltzmann-Enskog equation is for
small values of $q$ 
\begin{equation}
k_4\approx -2q +22.875 q^2 \label{k4series1}
\end{equation}

The fourth cumulant is measured in the MD simulations obtaining values 
consistent with the theoretical predictions.

The collision frequency $\nu$ is defined as the average number of
collisions per particle and unit $s$-time. So defined, $\nu$ is a
stationary quantity.
The simulation results fitted linearly with $q$ are
\begin{eqnarray}
\nu(n=0.05) &=& 0.189 - 0.0042 q \label{nuMD}\\
\nu(n=0.1)  &=& 0.403 - 0.022  q \nonumber \\
\nu(n=0.2)  &=& 0.929 - 0.277 q \nonumber
\end{eqnarray}

The collision frequency can be estimated using the the approximation
that there are no velocity correlations at contact (Eq.~\ref{nocorr}). 
Taking into account the distortion from the Maxwellian velocity
distribution, the collision frequency is given by 
\cite{noije98c} 
\begin{equation}
\nu = 2 \sqrt{\pi} n \chi \left(1 -\frac{k_4}{32}  \right) \label{nunocorr}
\end{equation}
where $\chi$ must be understood as the precollisional value.

Both $\chi$ and the term $(1-k_4/32)$ have a positive dependence on
$q$, but the MD results~(\ref{nuMD}) show a negative one. 
This discrepancy shows that the hypothesis of lack of velocity
correlations is false and must modified.
Also, when the numerical values for $\chi$ (Eq.\ref{chiMD}) and $\nu$
(Eq.~\ref{nuMD}) are replaced in the approximation (\ref{p1nocorr}) for
$p_1$ the predicted pressure is larger than the MD
value~(\ref{p1num}). That is, using the approximation that the two
particle distribution function can be factorized as in Eq. \ref{nocorr}
the predicted pressure is larger than the observed one.

\section{Velocity correlations at collisions}

Special MD measurements are done to study the source of the
discrepancies in pressure and collision frequency. We compute
numerically collisional averages sensible to the presence of velocity
correlations at contact for precollisional states. 

For a system composed of particles that interact with hard core
forces, the collision probability is given  in 2D by,

\begin{equation}
dP_{coll}(1,2) \propto 
\left|{\bf v}_{12}\cdot \hat{{\bf \sigma}} \right|
f^{(2)}(1,2) \delta({\bf r}_{12}-\hat{\bf \sigma})
\Theta\left(-{\bf v}_{12}\cdot{\bf r}_{12}\right)\,
d\hat{{\bf \sigma}}\, d{\bf v}_1\, d{\bf v}_2\, d{\bf r}_1\,d{\bf r}_2
\end{equation}
\noindent where $\sigma$ is the particle diameter
and $\Theta$ is the Heaviside function that restrict the velocities to
precollisional states. 
Collisional averages, defined as the average of any quantity at every
collision in the system, can be computed using the previous
probability distribution.
\begin{eqnarray}
\left< A \right>_{\mbox{coll}^{(-)}} &=& \frac{1}{n\nu} \int A(1,2) f^{(2)}(12)
 \left|{\bf v}_{12} \cdot \hat{{\bf \sigma}} \right| \Theta\left(-{\bf
  v}_{12}\cdot\hat{{\bf \sigma}}\right) d\hat{{\bf \sigma}}\,
d{\bf v}_1\, d{\bf v}_2 \\
&=& \frac{1}{n\nu} \int A(1,2) f^{(2)}(12)\left|{\bf v}_2 -{\bf v}_1
\right| db\, d{\bf v}_1\, d{\bf v}_2
\end{eqnarray}
where $b$ is the impact parameter. The sign $(-)$ in the average means
that $A$ is evaluated with the precollisional 
velocities and that ${\bf r}_{2}={\bf r}_1-\hat{\bf \sigma}$.
It must be remarked that the above defined collisional
average only takes into account the precollisional part of $f^{(2)}$.

In the hypothesis of absence of precollisional velocity correlations
at contact, the two particle  distribution function is written as in
Eq.~\ref{nocorr}. As the mean velocity vanishes, this approximation
implies that the following collisional average should vanish.
\begin{equation}
\Gamma = \left< \frac{{\bf v}_1\cdot{\bf v}_2}{\left|{\bf v}_2
-{\bf v}_1 \right|} \right>_{\mbox{coll}^{(-)}} 
\end{equation}
It must be remarked that in an elastic fluid at equilibrium
Eq.~\ref{nocorr} is exact and $\Gamma$ must vanish exactly.

\begin{figure}[htb]
\begin{center}
\epsfclipon
\epsfig{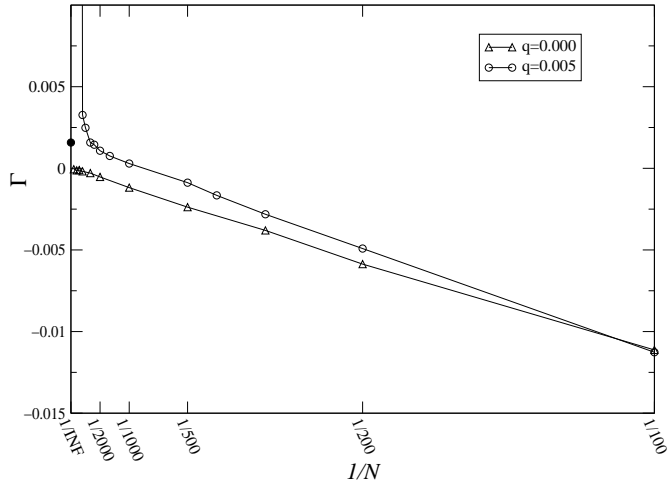}
\end{center}
\caption{Values of $\Gamma$ obtained in MD simulations for an elastic
case and a dissipative case as a function of the inverse of the number 
of particles $N$. The solid circle indicates the
extrapolated value for the dissipative system. The global density is $n=0.1$.}
\label{fig.prodcols}
\end{figure}

This property makes $\Gamma$ a good test for the hypothesis of the
lack of velocity correlations in stationary states.
This quantity couples strongly with the velocity fluctuation so careful
extrapolation to the infinite system must be done. In
Fig.~\ref{fig.prodcols} we show the extrapolation procedure for a
typical series of simulations. For small system sizes $\Gamma$ is
negative as a consequence of the conservation of total momentum.
In finite systems if one particle has a velocity $v_0$ the others have 
in average a velocity equal to $-v_0/(N-1)$, leading to negative
values for $\Gamma$. For large systems, the shearing instability
appears giving rise to a dramatic increase of $\Gamma$. To get the
extrapolated value to infinite system size, we consider systems up to
a certain size where no signal of the instability is present ($N=1500$ 
in the case of the figure).

After extrapolation to infinity system size, the obtained results
fitted linearly with $q$ are 
\begin{eqnarray}
\Gamma(n=0.05) &=& 0.097 q\\
\Gamma(n=0.1)  &=& 0.269 q \nonumber\\
\Gamma(n=0.2)  &=& 0.442  q \nonumber
\end{eqnarray}
That can be collected in the general expression
\begin{equation}
\Gamma = 2.29\, q\, n 
\end{equation}

This result means that in effect there are short range velocity
correlations with origin in the dissipative character of
the fluid. The fact that the correlation is positive means that
particles arrive at collisions 
with velocities more parallel than if there were no correlations.
This phenomenon can be understood in terms of recollisions
since, for the IHS model, the velocities of the particles after a collision
become more parallel than in the elastic case. This parallelization
has two effects: first it 
increases the probability of having a recollision (that is, after
colliding with a third particle the tagged pair recollides) and, second, when 
the pair recollides, their velocities are correlated, being more
parallel.
The $q$ dependence and density dependence can be well understood in
this model because the parallelization is proportional to $q$ and the
recollision probability to $n$.

This interpretation is also supported by the results in pressure and
collision frequency. In effect, the parallelization of velocities
after collisions produces that, in recollisions, the approaching
relative velocities 
are smaller, decreasing the collision frequency. Also the transfered
momentum --which, when averaged, gives $p_1$-- is smaller at
each collision. The 
combination of these two effects gives that the velocity correlations
produce smaller values for the pressure. The fact that the effects of
the correlations in $\Gamma$ and $p_1$ are numerically 
similar is also an indication that the two effects have the same origin.

At low density, precollision velocity correlations are small. This
explains the good agreement between the simulations and the theoretical
predictions using Enskog's theory for $\chi(\theta)$ in
Fig.~\ref{chi.fig}. 

Precollisional velocity correlations are also present in elastic
fluids, but only under nonequilibrium conditions. Low frequency and
long wavelength phenomena (where hydrodynamics is valid) are usually
studied as perturbations over local equilibrium states where no velocity
correlations exist. In granular fluids, the system is
always out of equilibrium and velocity correlations are always
present. For a given dissipation, there is no regime with no velocity
correlations around which perturbation analysis can be done. 

In elastic fluids, the factorization (\ref{nocorr}) is exact at equilibrium
and it allows to compute static properties like the pressure and the
collision frequency. From this starting point, Enskog equation is built
to describe the time evolution of the system in an approximate way (it
neglects precollisional velocity correlations in every regime). To mimic the
Enskog approach for granular fluids, we should take an equation that
includes velocity correlations, even in the HCS, in order to predict
accurately the static properties: pressure, collision frequency,
dissipation rate.

At low density and/or dissipation, the velocity correlations are small.
Then, the Boltzmann-Enskog equation can be used to describe granular
fluids with the same degree of approximation as for elastic
fluids. For dense dissipative systems more complex theories, that take into
account recollisions (for example, Ring Equations
\cite{Ring1,noije98}), are necessary. 

\section{Conclusions} \label{Conclusions}

Different properties for  the IHS model for granular fluids, put in the
homogeneous cooling state, have been carefully studied. Using a time
rescaling formalism it was 
possible to obtain precise averages in MD simulations, allowing to
study the dissipation and density dependence of these properties.
Dissipation took values from $q=0$ to $q=0.02$ and number density
from $n=0.05$ to $n=0.2$.

In all cases it was found that the velocity distribution is not
Maxwellian and the fourth cumulant $k_4$ is different from zero. Its
value does not depend on density and it is in good accord with the
kinetic theory predictions.

A distinction is made between the pair correlation function at contact 
for precollisional and postcollisional states. The first is the one
used in kinetic theory (Enskog's theory) and it was found that is has a
very small dependence on dissipation, taking the same value than for
elastic disks. The postcollisional pair correlation function takes larger
values and it can be fully predicted in terms of the precollisional
function.  

Collisional averages indicate that particles that are about to collide 
are correlated in a non trivial way, particles arrive to collisions
with velocities that are more parallel that in a elastic fluid. The
computed correlation, that is 
proportional to density and dissipation, has its origin in
recollisions: due to dissipation, particles that collide emerge with
more parallel velocities  than in the elastic case and, when they
recollide, their velocities are still more parallel.
Results obtained for pressure and collision frequency 
also show the signature of velocity correlations at collisions. The effect
of these correlations is to reduce the collision frequency and the
transfered momentum at collisions, thus reducing the virial pressure.

In elastic systems, velocity correlations are also present, but only
in nonequilibrium regimes. 
The intensity of the correlations reduces as the system approaches
equilibrium. 
In granular fluids, on the other hand, the dissipative
character of collisions puts the system always out of equilibrium,
creating velocity correlations. The observed
correlations are intrinsic to granular fluids since they are present
in every regine. There is no need for
special initial conditions or boundary conditions to obtain and compute
them. This allowed us to compute them in a very simple regime, the
homogeneous cooling state, with very high precision at the shortest
possible scale, the microscopic one. 

The presence of these correlations implies that the Enskog
factorization (\ref{nocorr}) is insufficient to compute the static
properties of the fluid: pressure, collision frequency, and dissipation
rate. For elastic fluids, Enskog's equation, even
if it is an approximation, accurately predicts static
properties. An equivalent approach for dissipative systems would need
the use of a kinetic theory that includes velocity correlations, even
in the HCS. More complex theories like ring kinetic theory
\cite{Ring1,noije98} or mode coupling theories \cite{PomeauResibois} are then
needed to describe dense granular fluids at finite dissipation.

\acknowledgments{We wish to thank J. Piasecki for useful discussions. We
also thank the referee for clarifying arguments and references. This work is
supported by a European Commission DG 12 Grant PSS*1045 and by a grant from
FNRS Belgium. One of us (R.S.) acknowledges the grant from {\em MIDEPLAN}.}


\appendix

\section{Relation between the post and precollisional part of
$\chi(\theta)$} \label{apendix.chi} 

In this appendix we will deduce the expression~(\ref{chipostpre}). The
deduction 
is based on the transformation of the distribution function at
collisions and in geometrical aspects of the collision rule.

The instantaneous character of binary collisions in the hard sphere (disk)
system, implies that the two particle distribution function can be
written as \cite{Lutsko}
\begin{equation}
f^{(2)}(1,2)\delta(r_{12}-\sigma) = \Theta(-{\bf r}_{12}\cdot{\bf v}_{12})
f^{(2)}_0(1,2)+ \frac{1}{\alpha^2}\Theta({\bf r}_{12}\cdot{\bf v}_{12})
\hat{b}^{*} f^{(2)}_0(1,2) 
\end{equation}

The first term is the precollisional distribution function $f^{(2)}_0$. The
second term represent the postcollisional distribution function,
written in terms of the precollisional one. The operator $\hat{b}^{*}$ has
the effect of replacing the velocities with the precollisional values,
and the factor $1/\alpha^2$ comes from the the change in relative 
velocity and the Jacobian of the transformation \cite{noije98}.

The postcollisional part of the pair distribution function is then
\begin{equation}
f^{(2)}(1,2)\Theta({\bf r}_{12}\cdot{\bf v}_{12})\delta(r_{12}-\sigma) = 
\alpha^{-2} f^{(2)}_0(1^*,2^*)\Theta({\bf r}_{12}\cdot{\bf
  v}_{12})\delta(r_{12}-\sigma) \label{fprefpost}
\end{equation}
where $1^*$ and $2^*$ represent the state of the particles with
precollision velocities.

\begin{figure}[htb]
\begin{center}
\epsfclipon
\epsfig{file=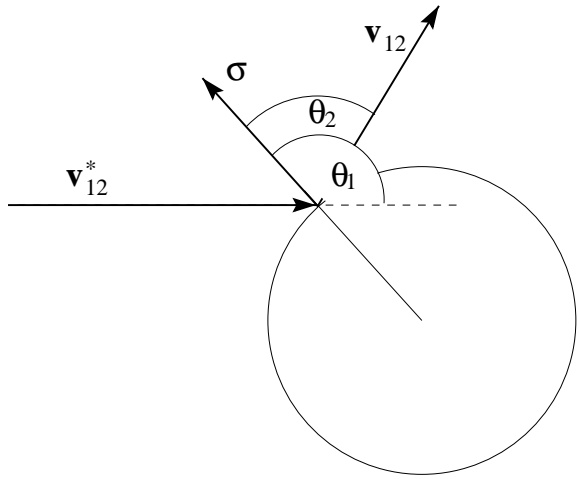,width=3in} 
\end{center}
\caption{Geometry of in inelastic collision. ${\bf v}^*_{12}$ (${\bf
v}_{12}$) is the incoming (outgoing) relative velocity and ${\bf \sigma}$
is the normal vector to the collision.}
\label{colision.fig}
\end{figure}

To simplify notation we will consider a collision in 2D; for the 3D
case, the analysis is similar and the results are summarized at the
end. The geometry of the collision
is represented in Fig.~\ref{colision.fig}. 
The postcollisional relative velocity is
${\bf v}_{12}$, the precollisional relative velocity is ${\bf
v}^*_{12}$, and $\sigma$ is the vector that joins the centers of the
two particles. The angles $\theta_1$ (precollision) and $\theta_2$
(postcollision) are defined as 
\begin{eqnarray}
\theta_1&=&\cos^{-1}\left( \frac{{\bf
      v}^*_{12}\cdot\sigma}{v^*_{12}}\right) \label{theta1}\\
\theta_2&=&\cos^{-1}\left( \frac{{\bf
      v}_{12}\cdot\sigma}{v_{12}}\right) \label{theta2}
\end{eqnarray}

The collision rule implies that 
\begin{equation}
\tan(\theta_2)= -\alpha^{-1} \tan(\theta_1)
\end{equation}
thus
\begin{equation}
\theta_1= \pi-\tan^{-1}\left(\alpha \tan(\theta_2)\right) \label{theta1theta2}
\end{equation}

The pair correlation function at contact $\chi(\theta)$ is defined as
\begin{equation}
\chi(\theta) =\frac{1}{n^2} \int f^{(2)}(1,2)\,\,
\delta\!\!\left[\theta - \cos^{-1} 
\left(\frac{{\bf v}_{12}\cdot\sigma}{v_{12}} \right) \right] d{\bf v}_1\,
d{\bf v}_2\, d\hat{\bf\sigma}
\end{equation}
where the factor $1/n^2$ guarantees the correct normalization.

For postcollisional angles, $f^{(2)}(1,2)$ can be expressed in terms of the
precollisional velocities using (\ref{fprefpost}). Changing integration
variables we obtain
\begin{equation}
\chi(\theta) =\frac{1}{\alpha n^2} \int f^{(2)}_0(1^*,2^*)\,\,
\delta\!\!\left[\theta 
 - \cos^{-1}\left(\frac{{\bf v}_{12}\cdot\sigma}{v_{12}} \right)
\right] d{\bf v}^*_1\, 
d{\bf v}^*_2\, d\hat{\bf\sigma} \qquad \theta<\pi/2
\end{equation}
where it has been used that $d{\bf v}_1\, d{\bf v}_2 = \alpha d{\bf v}^*_1\, 
d{\bf v}^*_2$~\cite{noije98}.

The argument of the delta function can be changed to precollisional
velocities using (\ref{theta1}), (\ref{theta2}),and
(\ref{theta1theta2})
\begin{eqnarray}
\chi(\theta) &=& \frac{1}{ n^2} 
\left[\cos^2(\theta)+\alpha^2\sin^2(\theta) \right]^{-1}  \nonumber \\
&&\int f^{(2)}_0(1^*,2^*)\,\,
\delta\!\!\left[-\tan^{-1}(\alpha \tan(\theta)) 
- \cos^{-1}\left(\frac{{\bf v}^*_{12}\cdot\sigma}{v^*_{12}} \right)
\right] d{\bf v}^*_1\, 
d{\bf v}^*_2\, d\hat{\bf\sigma} \qquad \theta<\pi/2
\end{eqnarray}
where the transformation rule for delta function has been used. The
integral can be identified as the pair correlation function at
contact for the precollisional angle $\left[\pi-\tan^{-1}(\alpha
 \tan(\theta))\right]$ 
\begin{equation}
\chi(\theta) = \left[\cos^2(\theta)+\alpha^2\sin^2(\theta)
\right]^{-1} \chi\left[\pi -\tan^{-1}(\alpha
 \tan(\theta) )\right] \qquad \theta<\pi/2
\end{equation}
that is, the postcollisional part of $\chi(\theta)$ can be computed
using the precolisional values.

In 3D, we define the pair correlation function that depends on the
solid angle $\hat{\Omega}$ that form the relative velocity and the
relative position, where $\hat{\Omega}$ is represented as
usual by the angles $\theta$ and $\phi$. As the tangential components of the
relative velocity are preserved at the collision, $\phi_1=\phi_2$. The
change on the normal component of the relative velocity implies the
relation (\ref{theta1theta2}). Using the generic relation of the delta
function $\delta(\hat{\Omega}-\hat{\Omega}')= \delta(\theta-\theta')
\delta(\phi-\phi')/|\sin(\theta)|$ and that $|\sin(\theta_1)| =
|\sin(\theta_2)|$, it is found by a similar analysis as in the 2D
case that
\begin{equation}
\chi(\hat{\Omega}) = \left[\cos^2(\theta)+\alpha^2\sin^2(\theta)
\right]^{-1} \chi(\hat{\Omega}^{*})  \qquad \theta<\pi/2
\end{equation}
where $\hat{\Omega}^{*}$ is the precollisional solid angle.

\end{document}